\documentclass[a4paper]{article}

\newcommand{\eps}{\varepsilon}
\newcommand{\nn}{\nonumber}
\renewcommand{\equiv}{:=}

\newtheorem{theorem}{Theorem}
\newtheorem{lemma}{Lemma}

\begin{document}

\title{Exponential Decay for Small Non-Linear Perturbations of Expanding Flat Homogeneous Cosmologies~\footnote{Work supported by CONICOR, CONICET, and SeCyT, UNC}}

\author{
 {\sc Oscar A. Reula}
  \thanks{Member of CONICET, email: reula@fis.uncor.edu}\\
  {\small FaMAF, Medina Allende y Haya de la Torre,}\\
  {\small Ciudad Universitaria, 5000 C\'ordoba, Argentina}\\
  {\small and} \\
  {\small Albert-Einstein-Institut} \\
  {\small Max-Planck-Institut f\"ur Gravitationsphysik} \\
  {\small Schlaatzweg 1, 14473 Potsdam, Germany} \\
}

\maketitle 

\vspace{-.4in}

\begin{abstract}
  It is shown that during expanding phases of flat homogeneous cosmologies all small enough
non-linear perturbations decay exponentially. 
This result holds for a large class of perfect fluid equations of state, but notably not for 
very ``stiff'' fluids as the pure radiation case.

\end{abstract}

\section{Introduction}
\label{sec:intro}

On expanding phases of nearly homogeneous cosmological models it is
believed that small perturbations are pulled apart and so washed away.
Thus, if they were the only type of perturbations present they would
decay to zero, making the model more and more homogeneous as time
elapses.  This intuitive picture of homogenization, basic in all
arguments used on inflationary models, has so far not been rigorously
justified beyond the corresponding one on the linearized equations off
a homogeneous background, the corresponding nonlinear case of
Newtonian cosmological models~\cite{BrauerU:CosNH}, or the
Electro-Yang-Mills-Vacuum case using conformal methods,
\cite{Friedrich:GlobE}.

The main mathematical difficulty to tackle this problem has been the
absence of a set of variables in which Einstein's equations coupled to
a fluid could at the same time have a well posed formulation and a
linearization off a homogeneous expanding background where all
eigenvalues have negative real part, for in that case, as we shall
argue in \S{}\ref{sec:energy}, decay of small (but nonlinear)
perturbations can be shown to follow quite easily. As we shall show in
\S{}\ref{sec:evol} this problem has been recently overcome by using a
novel set of \textsl{Lagrangian variables}, in which the time
component of a frame variable follows the fluid four-vector under
evolution, \cite{Friedrich:EvoEGF}.  Thus, using these variables we
are able to show exponential decay of small non-linear perturbations
for a wide range of equation of state. 

In the next section we introduce Friedrich's system for the
Einstein-Euler system of equations in Lagrangian variables.  Since
this system has a well posed initial value problem (it is symmetric
hyperbolic) we know local in time solutions exist, and so the only
question is whether they decay or not if they are sufficiently close
to an homogeneous expanding cosmology.  To answer this we look at the
linearization of this system off a homogeneous cosmological model, and
display the conditions under which the real part of the eigenvalues of
the system are negative definite. For the flat case, the only one
considered in detail, they turn out to be only conditions on the
equation of state, and the only one which seems to be a real
restriction within the class of equations of state usually considered
is condition~(\ref{cond_nu}), which limits the sound speed of the
fluid to be less than one third of the speed of light. This rules out
from consideration the stiff or pure radiation perfect fluids. As
noted by several colleagues, perhaps this limiting case could be
treated using the conformal Einstein's
equations~\cite{Friedrich:GlobE}.  
It seems that one can not go beyond
that limit with the present approach and thus one is led to wonder
whether this limitation has a physical origin, in which case it should
be explored in more detail.

In \S{}\ref{sec:energy} we complete the assertion on decay by giving a
general proof of exponential decay for symmetric hyperbolic systems
whose eigenvalues have negative definite real part, that is for system
as the one we are considering. This theorem is a simple adaptation to
the case under consideration of a more general class of results which
are known in the literature on non-linear decay, for a resent review
of them see~\cite{KreissH-O:StabTD}.

We present the results for flat backgrounds, for non-flat ones similar
results should hold as long as the expansion is large enough, but
further work is needed to fully understand how big this expansion has
to be in terms of invariant quantities, in particular how negative the
cosmological constant has to be, so we don't elaborate in detail here.


\section{The evolution equations}
\label{sec:evol}


We use the evolution system for the Einstein-Euler equations
as defined in \cite{Friedrich:EvoEGF}.  In that system one introduces a
set of orthonormal frame vectors $\{u^a, e^a_i \}$, $i=1..3$, $u^a u^b
g_{ab} = 1, e^a_i e^b_j g_{ab} = -\delta_{ij}$.  In accordance with
the Lagrangian coordinate scheme, the time-like component of the frame
is chosen to be the four-velocity of the fluid, (so in general it is not
surface orthogonal), and the other frame vectors are chosen to be
dragged along the integral curves of $u^a$, that is along the flow
lines, using Fermi transport, that is, $\Gamma_0{}^i{}_k = 0$. 
The remaining derivatives of the frame vectors are grouped into two set of
scalars, 
$a^i \equiv \Gamma_0{}^i{}_0$, 
and 
$\chi_{ij} \equiv -\Gamma_i{}^0{}_j$, 
they shall form part of the evolution system. 
Note that 
$\chi_{ij}$ 
is not symmetric, for $u^a$ is in general not surface orthogonal.  

Part of the fluid equations are implied by 
the evolution equations for the components of the
acceleration, $a^i$, and a constraint which we shall indicate latter.
The rest of them, and the only one surviving as such is the conservation equation for the energy density 
$\rho \equiv T_{ab}u^a u^b$,
equation (\ref{eqn:rho}).
 
To complete the evolution system one adds part of the
Bianchi identities to obtain evolution equations for the components of
the electric and magnetic decompositions of the Weyl tensor.
The resulting system is,

\begin{eqnarray}
  \label{eq:evol}
  \partial_t e^{\mu}_i & = & - \chi_i{}^j e^{\mu}_j - a_i e^{\mu}_0  \\
  \partial_t \Gamma_{ik}^j & = & - \Gamma_{lk}^j\chi_i{}^l - 2\chi_{i[l} a_{k]} h^{jl} - B_{ip}\eps^{pj}{}_k \\ 
  \partial_t a_i + \nu^2 D_k \chi_i{}^k & = & -(\chi_i{}^j + \beta \chi \delta_i{}^j + \nu^2 (\chi^j{}_i - \chi_i{}^j)) a_j \label{eqn:ace} \\ 
  \partial_t \chi_{ij} - D_j a_i & = & - \chi_i{}^l \chi_{lj} - a_i a_j + 2\nu^2 \chi \; \chi_{[ij]} \\
                       & &  - E_{ij} - \frac{\kappa}{2}(\frac{\rho}{3} + p) h_{ij} - \frac{\Lambda}{3} h_{ij} \nn \label{eqn:chi} \\
  \partial_t \rho & = & -(\rho +p) \chi \label{eqn:rho} \\
  \partial_t E_{ij} + D_k B_{l(i} \eps_{j)}{}^{kl} & = &  \chi_{(i}{}^k E_{j)k} + 2\chi^k{}_{(i} E_{j)k} - h_{ij} \chi^{kl} E_{kl} - 2\chi E_{ij}  \\          
                                                   &   & + 2 a_l \eps^{lk}{}_{(i} B_{j)k}  
                       - \frac{\kappa}{2} (\rho + p)(\chi_{(ij)} - \frac{1}{3}\chi h_{ij}) \nn \\
  \partial_t B_{ij} - D_k E_{l(i} \eps_{j)}{}^{kl} & = & \chi^k{}_{(i} B_{j)k} - \chi B_{ij} 
                                                         - \chi_{kl} B_{pq} \eps^{pl}{}_{(i} \eps^{qk}{}_{l)} \\          
                                                   &   & - 2 a_l \eps^{lk}{}_{(i} E_{j)k}  \nn  
\end{eqnarray}
where

\begin{equation}
  \label{eq:beta}
  \beta :=  \frac{\rho+p}{\nu^2}\frac{\partial^2 p}{\partial \rho^2} - \nu^2.
\end{equation}

This system is symmetric hyperbolic when the speed on sound ($\nu^2$)
is positive,
and so in that case it has a well posed initial value formulation. Thus there is an energy estimate and
local in time solutions exist and are unique for smooth enough initial
data sets in $L^2$. In what follows we shall use implicitly the
existence and uniqueness of these solutions.

For simplicity we are considering here only homentropic fluids.
In the more general case one needs further evolution equation for the entropy and
its space derivatives, see \cite{Friedrich:EvoEGF}. 
The evolution equation for the entropy just asserts it is conserved along integral curves of
the fluid four-vector, the evolution equation for the space derivatives of the entropy, assert
consequently that they decay at the precise rate at which the expansion smoothes things out.
Since these evolution equations are not coupled at the principal part level,
one can always treat them as we shall treat the frame in \S\ref{subsec:variant} and obtain decay properties
without having to impose further conditions on the system.

In addition to these evolution equations there are a number of constraint equations, the usual
ones that arise in general relativity, plus some connecting the different fields, for instance
those which guarantee that the $\chi_{ij}$ are part of the connection coefficients of the 
frame. Here we give a brief discussion on the constraint problem, but do not go in details,
for the aim of this work is to study the time behavior of solutions once initial data for them
--satisfying the constraint-- are given. 

To solve the constraints one would follow the usual procedure for
frame equations, and so we only remark on the differences which appear
in this case.  One first solve for a three metric and extrinsic
curvature the usual scalar and vector constraints with some boundary
or asymptotic conditions.  Here we consider either an initial surface
whose background is a flat 3-torus, or a flat $R^3$.
In the case of a 3-torus we consider periodic
boundary conditions~\footnote{Here the perturbation of the mass
density has to be adjusted so that there can be solutions.}. 
In the open case ($R^3$) we consider isolated
perturbations in the sense that one requires the perturbations to
decay to zero asymptotically and are of finite energy. In the general
case the perturbations in $R^3$ decay too slowly for the frame perturbations
to be in $L^2$, for there is a mass perturbation, and so one must use
the variant of the decay theorem discussed in \S\ref{subsec:variant}.
One can also tune the mass density perturbation so that the total mass
perturbation vanishes, but these seems to restrict the space of
allowed perturbations. For this subclass of perturbations one can use the
standard decay theorems given in \S\ref{sec:energy}.  Once this is
done one then provides a lapse-shift pair which in this case can be
thought of as the 3+1 decomposition of the fluid four vector.  Thus
one has a four metric, and one knows, using the field equations, all
space and time derivatives of that metric at the initial surface.  One
then chooses a frame for that metric so that the time-like component
of that frame coincides with the fluid four-vector. Most of the time
derivatives of that frame are fixed from the metric condition and so
are given by the extrinsic curvature of the surface, the rest,
amounting to frame rotations and accelerations are fixed by the gauge
condition imposed (Fermi transport) on the frame and from the
acceleration, which is given by the Euler's equation,
(\ref{eqn:euler}), which here is part of the constraints.  Thus we
have all derivatives of the frame at the initial surface and so can
compute the frame coefficients and from them all frame related
quantities which enter as dynamical variables.  They satisfy
automatically the rest of the constraints.  In the same way one can
compute the Weyl tensor at the surface and its frame components, and
so complete the set of initial data.

One does not need for this case to prove that the constraints are
propagated for this follows from general arguments, see for
instance~\cite{ReulaOA:HypME}. Indeed suppose initial data is given
satisfying the constraint, and suppose for contradiction that at some
point $p$ inside the domain of dependence of the initial slice the
constraints cease to be satisfied. Then we can take a small
neighborhood of that point and inside it, to the past of it, construct
a space-like slice so that $p$ is in its domain of dependence.  In
that slice the smooth induced initial data satisfies the constraints
and so we can generate a local solution of Einstein's equation using,
say, the harmonic gauge. That solution satisfies the constraints along
evolution. Furthermore we can construct locally a frame whose
time-like component points along the fluid and whose other components
are propagated using Fermi transport. The proof of this involves only
the theory of ordinary differential equations.  Thus we have a
solution of Einstein's equations in the gauge of our propagation
equations and so it is a solution to them, furthermore all constraints
are satisfied, even at $p$.  But the solutions to the evolution
equations are unique, and so we reach a contradiction, implying the
constraints must propagate correctly inside the domain of dependence.
The constraint propagation has also been checked explicitly, \cite{Friedrich:PC}

In the above system it appears $\nu^2$ in the denominator, so it is
not immediate that this equations behave nicely in the limit when this
quantity --the sound speed-- goes to zero, and for many equations of
state this happens when the density goes to zero.  Here we are not
interested in reaching the limit where the expansion makes the density
go to zero, but nevertheless it is interesting to obtain bounds which
are uniform in $\nu^2$ in the whole region where it becomes small. An
easy way to do this is to rescale the acceleration with a factor $\nu$
and so get a system whose coefficients are bounded even in the limit
$\rho \to 0$. Defining $\tilde{a}_i = \frac{1}{\nu} a_i$, and
multiplying equation (\ref{eqn:ace}) by $\frac{1}{\nu}$ we get,

\begin{eqnarray}
  \label{eq:evol_tilde}
  \partial_t e^{\mu}_i & = & - \chi_i{}^j e^{\mu}_j - \nu \tilde{a}_i e^{\mu}_0  \\
  \partial_t \Gamma_{ik}^j & = & - \Gamma_{lk}^j\chi_i{}^l 
                             - 2 \nu \chi_{i[l} \tilde{a}_{k]} h^{jl} - B_{ip}\eps^{pj}{}_k \\ 
  \partial_t \tilde{a}_i + \nu D_k \chi_i{}^k & = & 
                             - (\chi_i{}^j + \tilde{\beta} \chi \delta_i{}^j 
                             + \nu^2 (\chi^j{}_i - \chi_i{}^j)) \tilde{a}_j \\
  \partial_t \chi_{ij} - \nu D_j \tilde{a}_i & = & - \chi_i{}^l \chi_{lj} 
                       - (\gamma +\nu^2)\tilde{a}_i \tilde{a}_j 
                       + 2\nu^2 \chi \; \chi_{[ij]} \\
                       & &  - E_{ij} - \frac{\kappa}{2}(\frac{\rho}{3} + p) h_{ij} 
                       - \frac{\Lambda}{3} h_{ij} \nn \\
  \partial_t \rho & = & -(\rho +p) \chi \\
\partial_t E_{ij} + D_k B_{l(i} \eps_{j)}{}^{kl} & = &  
                   \chi_{(i}{}^k E_{j)k} + 2\chi^k{}_{(i} E_{j)k} 
                   - h_{ij} \chi^{kl} E_{kl} - 2\chi E_{ij}  \\          
                   &   & + 2 \nu \tilde{a}_l \eps^{lk}{}_{(i} B_{j)k}  
                   - \frac{\kappa}{2} (\rho + p)(\chi_{(ij)} 
                   - \frac{1}{3}\chi h_{ij}) \nn \\
  \partial_t B_{ij} - D_k E_{l(i} \eps_{j)}{}^{kl} & = & 
                     \chi^k{}_{(i} B_{j)k} - \chi B_{ij} 
                     - \chi_{kl} B_{pq} \eps^{pl}{}_{(i} \eps^{qk}{}_{l)} \\          
                     &   & - 2 \nu \tilde{a}_l \eps^{lk}{}_{(i} E_{j)k}  \nn  
\end{eqnarray}
where $\gamma(\rho) \equiv - \frac{(\rho + p)}{2 \nu^2} \frac{d^2p}{d\rho^2}$
and $\tilde{\beta} \equiv \beta - \gamma = -3\gamma - \nu^2$,
to get these equations we have used the evolution equation for $\rho$, (\ref{eqn:rho}), and one
of the constraint equations for the system, namely one of the fluid equations in the 
Euler description,
\begin{equation}
  \label{eqn:euler}
  D_j p = (\rho + p) a_j
\end{equation}

This system is symmetric hyperbolic even for $\nu = 0$, and so well
adapted for the case in which the mass density tends to zero. But to
have a well posed system one also needs that the coefficient of the
system be smooth.  It can be seen that for many realistic equation of
state $\gamma$ is bounded as a function of $\rho$, but in general it
is not differentiable. For instance, for an equation of state of the
form $p = k\rho^{\tilde{\gamma}}$, $1 \leq \tilde{\gamma} \leq
\frac{4}{3}$ second derivatives are not bounded when $\rho \to 0$, 
unless the relation between $p$ and $\rho$ becomes
linear.~\footnote{For an equation of state of the form $p =
  k\rho^{\tilde{\gamma}}$ one has, $\gamma(\rho) =
  \frac{(\tilde{\gamma} - 1)}{2}(1 + k \rho^{\tilde{\gamma} -1})$} 
But
notice that using the evolution equation for $\rho$ one can see that
time derivatives of $\gamma$ are bounded [for there is a contribution
of an extra factor $(\rho + p)$].  Similarly, using (\ref{eqn:euler}),
one can see that the space derivatives of $\gamma$ are also
bounded~\footnote{For the case $p = k\rho^{\tilde{\gamma}}$ one can
  see that all derivatives are polynomial expressions in
  $\rho^{\frac{\tilde{\gamma} - 1}{2}}$, and so regular, as long as
  $\tilde{\gamma} \geq 1$.}.  So, in general one can not proceed to
make a priori estimates for derivatives of the fields, --as needed for
showing existence-- for equation (\ref{eqn:euler}) is known only a 
posteriori to hold.  
But
nevertheless one can make a posteriori estimates on the solution, for
if we know that the constraint are satisfied we can, using
(\ref{eqn:euler}), obtain bounds on the space derivatives of the
solutions to any desired order. This shall be enough for our purposes
of obtaining uniform (in $\nu^2$) estimates in regions where solutions
are known to exist.


\subsection{The background solution}

\label{subsec:back}


We take as background solution a Friedmann--Robertson-Walker Universe, namely,

\begin{equation}
\label{eqn:metric}
ds^2 = dt^2 - \frac{a_0(t)^2}{\omega^2}(dx^2+dy^2+dz^2), 
\end{equation}
with $\omega = 1 + kr^2/4$, and $a_0(t)$ solution of,
\begin{equation}
  \label{eq:a(t)}
  (\frac{\dot a_0}{a_0})^2 + \frac{k}{a_0^2} - \frac{\kappa \rho_0}{3} + \frac{\Lambda}{3} = 0,
\end{equation}
where $k=1,-1,0$ determines the curvature of the surface, $\Lambda \leq 0$ is the
cosmological constant, and $\rho_0$ is the background energy density.

The background density satisfies,

\begin{equation}
  \label{eq:rhocero}
  \dot \rho_0 = -3h_0 (\rho_0+p_0),
\end{equation}
where 
$h_0 = \frac{\dot a_0}{a_0}$, 
is the Hubble function and 
$p_0=p_0(\rho_0)$ 
is the pressure, which is assumed to be a function of the
density only. We shall assume that either $\Lambda < 0$ or that we are in
a time period where $h_0 > \bar{h}_0 > 0$, for some constant $\bar{h}_0$.

It is easy to check that all evolution and gauge equations are satisfied
if in the above coordinates one takes the following frame:

\begin{equation}
  \label{eq:backframe}
  {}_0u^{\mu} = {}_0e_0^{\mu} = (1,0,0,0), \;\;\;\;\;\; {}_0e_i^{\mu} = (0,\frac{\omega}{a_0}\delta_i^{\mu})
\end{equation}
The corresponding fields for this frame take the form:

\begin{eqnarray}
  \label{eq:connection}
  {}_0\Gamma^i{}_{jk} &=& \frac{k\omega}{2a_0^2}(-h_{jk}x^i + \delta^i_j x_k), \\
  {}_0\Gamma_i{}^0{}_j &=& -{}_0\chi_{ij} = -h_0 \; h_{ij} \\
  \mbox{all others}  &=& 0.
\end{eqnarray}
where $x^i = {}_0e^i_{\nu} x^{\nu} = -\frac{\omega}{a_0}\delta^i_{\nu}x^{\nu}$.

The proper time of co-moving observers would be taken to be the time with respect to which we
assert the exponential decay result. Given any perturbed space-time this time is no longer
an invariant, but it is so up to order $\eps$, that is up to the size of the perturbation.
Furthermore, while the expansion is producing an exponential decay the difference between this
time and the proper time of geodesic observers is uniformly bounded and goes to zero as $\eps \to 0$.


\subsection{The linearized equations}
\label{subsec:linear}


The linearized equations are, 

\begin{eqnarray}
\label{eq:Linear}
\dot{ {}_1e^{\mu}_{i}} &=& -h_0 \; {}_1e^{\mu}_{i} - {}_1\chi_i{}^j \; {}_0e^{\mu}_j  
                           -  \; \nu_0 {}_1\tilde{a}_{i} \delta^{\mu}_0 \\
  \dot{ {}_1\Gamma^i_{jk}} &=& -h_0 \; {}_1\Gamma^i_{jk} - \eps^{li}{}_{k} \; {}_1B_{jl} \\
\dot{ \; {}_1\tilde{a}_{i}} + \nu_0 \; D_j {}_1\chi_i{}^j &=& - h_0 \alpha \; {}_1\tilde{a}_{i} \\
\dot{ {}_1\chi_{ij}} - \nu_0 \; D_j {}_1\tilde{a}_{i} &=& 
                           -2h_0 ({}_1\chi_{ij} + 3\nu^2_0 \; {}_1\chi_{[ij]}) 
                           -  {}_1E_{ij} \\
                           & & - \frac{\kappa}{6}(1-3\nu_0^2) h_{ij} \rho_1  \nn \\
\dot{ {}_1B_{ij}} - D_k({}_1E_{m(i})\eps_{j)}{}^{km} &=& -3h_0 \; {}_1B_{ij} \\
\dot{ {}_1E_{ij}} + D_k({}_1B_{m(i})\eps_{0j)}{}^{km} &=& -3h_0 \; {}_1E_{ij} - 
              \frac{\kappa}{2}(\rho_0+p_0)({}_1\chi_{ij}-\frac{1}{3} \; {}_1\chi h_{ij}) \\
\dot \rho_1 &=& -3h_0(1+\nu^2_0) \rho_1 - (\rho_0+p_0) \; {}_1\chi 
\end{eqnarray}
where

\begin{eqnarray}
  \label{eq:alpha}
  \alpha & = & (1 - 3\nu^2 + 9\frac{\rho+p}{2\nu^2}\frac{\partial^2 p}{\partial \rho^2}) \\
         & = & 1 - 3\nu_0^2 - 9\gamma, \nn
\end{eqnarray}


\subsection{Negativity of the eigenvalues}
\label{subsec:neg}


We notice that our system has the following form,

\begin{equation}
  \label{eq:syst}
  u_t = A^a(u)D_a u + (B_0 + \eps B_1(u))u,
\end{equation}
where the matrix $A^a(u)$ is symmetric, in the sense that there exists a strictly positive symmetric 
bilinear form $H(u)$, smooth in $u$,  such that
$H(u)(A^a(u)) - (A^a(u))^{\dagger}H(u) = 0$.

Due to this property it follows quite easily that the eigenvalues of the system have strictly negative real part
for small enough $\eps$ if and only if $HB_0 + B^{\dagger}_0H$ is a negative definite bilinear form.
In the next section we shall show that the strict negativity of the eigenvalues suffices to show exponential decay.

It can be easily seen that a family of bilinear forms making the
system symmetric-hyperbolic is given by,

\begin{eqnarray}
  \label{eq:H0}
  <(e,\Gamma,a,\chi,E,B,\rho),&& \!\!\!\!\!\!\!\!\!\! H_0 (e,\Gamma,a,\chi,E,B,\rho)>
\!\!\!\!\!\!\!\!\!\!\!\!\!\!\!\!\!\!\!\!   \\
                = & & + C_e^2 {}_0l_{\mu \nu} h^{ij} e^{\mu}_i e^{\nu}_j \nn \\
                  & &  - C_{\Gamma}^2 h_{ij} h^{kl} h^{mn} \Gamma^i_{km} \Gamma^j_{ln}                  
                     - C_{\chi}^2 h^{ij} \tilde{a}_i \tilde{a}_j             \nn  \\
                  & &   +  C_{\chi}^2[h^{im} h^{jn} (\chi^{ST}_{ij}\chi^{ST}_{mn}
                     + \chi^{A}_{ij}\chi^{A}_{mn}) + \frac{1}{9} \chi^2] \nn \\
                  & & + h^{im} h^{jn}E_{ij}E_{mn} + h^{im} h^{jn}B_{ij}B_{mn} + C_{\rho}^2 \rho^2 \nn
\end{eqnarray}
where $\chi^{ST}_{ij}$ ($\chi^{A}_{ij}$) denote the symmetric trace
free, and the antisymmetric parts of $\chi_{ij}$ respectively, 
$l_{\mu\nu} \equiv {}_0h_{\mu\nu} -2\; {}_0 \; u_{\mu} {}_0u_{\nu}$, 
and $C_e$, $C_{\Gamma}$, $C_{\chi}$, and $C_{\rho}$ are constants 
which can take any nonzero value, reflecting the fact that the
bilinear form which symmetrizes the system is not unique. 
These constants shall be determined 
bellow in such a way as to maximize the range on which the negativity
of $B_0$ holds. Notice that $H(u) = H_0$, a bilinear form which only
depends on the background solution and so it is constant in space
directions. 

Using the values of $B_0$ from the linearized equations we obtain, 

\begin{eqnarray}
  \label{eq:ineqB}
  \frac{1}{2h_0}<(e,\Gamma,a,&&\!\!\!\!\!\!\!\!\!\!\chi,E,B,\rho),(H_0B_0 + B^{\dagger}_0H_0) (e,\Gamma,a,\chi,E,B,\rho)> \!\!\!\!\!\!\!\!\!\!\!\!\!\!\! \\
          =& & - C_e^2 {}_0l_{\mu \nu} h^{ij} e^{\mu}_i e^{\nu}_j 
                  + C_{\Gamma}^2 h_{ij} h^{kl} h^{mn} \Gamma^i_{km}
             \Gamma^j_{ln} \nn \\
          & & + \alpha C_{\chi}^2  h^{ij} \tilde{a}_i \tilde{a}_j 
           - 2C^2_{\chi}[(1-3\nu_0^2)(\chi^A)^2 + (\chi^{ST})^2 
                         + \frac{1}{9}\chi^2] \nn \\
          & & - 3(E)^2 -3 (B)^2 - 3C_{\rho}^2(1+\nu_0^2) \rho^2  \nn \\
             && - \frac{\nu_0}{h_0} C_{e}^2 {}_0l_{\mu \nu} h^{ij} \tilde{a}_i e^{\nu}_j \delta^{\mu}_0 
                - \frac{1}{h_0} C_{e}^2 {}_0l_{\mu \nu} h^{ik} \chi_i{}^j e^{\mu}_j e^{\nu}_k \nn \\
             && - \frac{1}{h_0} C_{\Gamma}^2 B_{lj} \eps^{li}{}_k h_{ip} h^{jm} h^{kn} \Gamma^p_{mn} \nn \\
             && - \frac{1}{h_0} [C_{\chi}^2 + \frac{\kappa}{2}(\rho_0+p_0)]E\chi^{ST} \nn \\
             && - \frac{1}{h_0} [\frac{\kappa}{6}C_{\chi}^2(1-3\nu_0^2) + C_{\rho}^2(\rho_0+p_0)]\rho\chi \nn
\end{eqnarray}
Thus negativity of the eigenvalues follows if we can choose the constants on the diagonal of $H_0$ so that:

\begin{eqnarray}
  \label{eq:cond.}
 \frac{1}{4} \frac{C_e^2}{\alpha^2 C_{\chi}^2} &<& h_0^2 \\
  \frac{1}{4} \frac{\nu_0^2 C_e^2}{\alpha^2 C_{\chi}^2} &<& h_0^2 \\
  \frac{1}{4} C_{\Gamma}^2 &<& h_0^2 \\
  3\nu_0^2 - 9 \frac{\rho_0 + p_0}{2\nu_0^2} \frac{\partial^2 p}{\partial \rho^2} &<& 1 \;\;\;\;\;\;\; (\alpha > 0) \label{cond_alpha} \\
  3\nu_0^2 &<& 1 \label{cond_nu} \\
  \frac{1}{2\sqrt{6}}[C_{\chi} + \frac{\kappa(\rho_0+p_0)}{2 C_{\chi}}] &<& h_0  \\
  \frac{1}{\sqrt{2}\sqrt{1+\nu_0^2}}[\frac{\kappa (1++3\nu_0^2) C_{\chi}}{6 C_{\rho}} + \frac{(\rho_0+p_0) C_{\rho}}{C_{\chi}}] &<& h_0 
\end{eqnarray}

We now choose the constants in such a way to obtain the desired
bounds.  
The first and the second can always be satisfied by choosing $C_e$ small enough.
In fact, from the arguments used for the case where the frame perturbations
are not square integrable one sees that these to conditions are superfluous.
Note that the second is implied by the first if the fifth inequality is taken into
account.
The third can be always satisfied by choosing $C_{\Gamma}$ small
enough.
The fourth inequality is a real constraint on the equation of
state of the fluid, as is the fifth, which in particular rules out
from our considerations pure radiation Universes, both are very
similar.  
The sixth inequality can be maximized choosing $C_{\chi}^2 = \kappa
(\rho_0 + p_0) / 2$, and the seventh, choosing $C_{\rho} =
\frac{\kappa}{\sqrt{24}}$. Both give similar conditions, the first
gives,
\begin{equation}
  \label{eq:cond.5}
  \kappa (\rho_0 + p_0) < 12 h_0^2,
\end{equation}
while the second, 
\begin{equation}
  \label{eq:cond.6}
  \kappa (\rho_0 + p_0) < 12\frac{(1+ \nu_0^2)}{1+3\nu_0^2} h_0^2 \leq 12 h_0^2
\end{equation}
and so the second implies the first, and both are satisfied if (\ref{cond_nu}) holds
and 
\begin{equation}
  \label{eq:cond.6bis}
  \kappa (\rho_0 + p_0) < 8 h_0^2
\end{equation}

Using now (\ref{eq:a(t)}), and again (\ref{cond_nu}) to estimate $p_0 < \rho_0/3$ in the second 
condition we get,

\begin{equation}
  \label{eq:cond.c}
  \frac{k}{a(t)^2} < (\frac{\kappa \rho_0}{6} - \frac{\Lambda}{3}),
\end{equation}

Since the right hand side is positive, while $k=-1,0,1$,  we see that this is only a condition
for the case of closed cosmologies, namely $k=1$. In that case it says,

\begin{equation}
  \label{eq:cond.d}
  1=k < \frac{1}{2}(\frac{2M_0}{R_0} - \frac{2}{3}\Lambda R_0^2)
\end{equation}
where $M_0$, and $R_0$ are the mass and radius of the Universe. Thus slowly expanding Universes
with mass over radius close to closure and small cosmological constant can not be treated with our method.

We summarize this in the following Lemma:

\begin{lemma}

If the equation of state is such that conditions (\ref{cond_alpha}--\ref{cond_nu}) are satisfied, and
for the case $k=1$ condition (\ref{eq:cond.d}) is also satisfied, then $H_0B_0 + B_0^{\dagger}H_0$ is
negative definite.

\end{lemma}


\section{The Energy Argument}
\label{sec:energy}


In this section we recall the standard argument leading to asymptotic stability of solutions from
conditions on the non-principal part of a symmetric hyperbolic system.

We begin considering a system of the form:

\begin{equation}
  \label{eq:syst0}
  u_t = (A^a_0 + \eps A^a_1(u))D_a u + (B_0 + \eps B_1(u))u
\end{equation}

with:

\begin{enumerate}

\item The matrices $A^a_0$, and $B_0$ are constant in space and time directions.

\item Symmetric Hyperbolicity: there exists a strictly positive symmetric bilinear form 
$H = H(u) = H_0 + \eps H_1(u)$
smooth in $u$, such that \\
$H(u)(A^a_0 + \eps A^a_1(u)) - (A^a_0 + \eps A^a_1(u))^{\dagger}H(u) = 0$.

\item Decay condition: The matrix $B_0$ is negative definite, namely \\
$<u,(H_0 B_0 + B_0^{\dagger}H_0)u> \leq -\delta<u,H_0u>$, 
for some $\delta > 0$.

\end{enumerate}

Remarks:

\begin{enumerate}
\item Perturbations of flat Friedmann-Robertson-Walker space-times can
  almost be cast in this form by splitting the solution as a
  background solution plus a small part, $u_E := u_0 + \eps u$, and 
  noticing that the background solution (out of which one constructs 
  $A^a_0$, and $B_0$) is constant in space directions.
  In this case, the background solution, as well as $A_0^a$, $B_0$,
  and $H_0$ depend on time, and so we shall latter modify the present
  argument accordingly.

\item For our system $H=H_0$, for it can be chosen only to depend on
  the background fields, and on the tetrad components of the metric,
  which by definition are constant (in space) scalars.

\item For our system, and only for convenience in notation, we shall
  later use derivatives along the frame directions and not $D_a$.

\item The topology of the initial space can be either compact, $T^3$,
  say, or open $R^3$.
  In the presence of constraints the open case might be uninteresting
  for there could be too few solutions of the constraints which are in
  the required function spaces for showing existence.
  We shall make latter a variant of the standard theorems which
  requires less stringent conditions on the initial data. 
  We shall always assume we have solutions of the constraint equations
  in the required functional spaces.

\end{enumerate}

We now define an {\sl energy vector} by:

\begin{equation}
  \label{eq:energy_vector}
  E^a := (\partial_t)^a <u,Hu> + <u,HA^a u>,
\end{equation}
and compute,

\begin{eqnarray}
  \label{eq:div}
  \nabla_a E^a 
         & = & <u_t,Hu> + <u,H u_t> + <u,H_t u> \\
         & + & <D_a u, HA^a u> + <u,HA^aD_a u> + <u,D_a(HA^a)u> \nn \\
         & = & 2\Re <u,H(u_t + A^aD_au)> + <u,(H_t + D_a(HA^a))u> \nn \\
         & = & 2\Re <u,H_0 B_0u>  + \eps [2\Re <u,(H_0+\eps H_1) B_1u> \nn \\
         & + & <u,((H_1)_t + D_a(H_0A^a_1+H_1 A^a_0 + \eps H_1A^a_1))u>] \nn 
\end{eqnarray}

Similar expressions hold for space derivatives of $u$, (i.e. tangent to the family of hypersurfaces) 
since due to the constancy along space and time directions of $A^a_0$ and $B_0$, 
they satisfy similar equations with same zeroth order (in $\eps$) terms.
Thus, considering a $E^a_p$ sum of all the $E^a$'s for $u$ and all its
derivatives up to order 
$p \geq 3$, 
integrating in space and time its divergence, using Gauss theorem, and
(for simplicity) taking the limit when the integration region shrinks
to zero on the time direction we get,

\begin{equation}
  \label{eq:energy_ineq}
  \frac{d}{dt} (E_p) \leq -\frac{\delta}{2}\;E_p + \eps F(E_p),
\end{equation}
where

\begin{equation}
E_p(u) := \int_{\Sigma_{\tau}} E_p^a n_a(u) \; d\Sigma,
\end{equation}
$n_a$ is the normal to a family of space-like hypersurfaces
parametrized  with $\tau$, and where we have used standard Sobolev
and Gagliardo-Niremberg-Moser estimates to control all the nonlinear terms.

Thus, given an initial value for $u$ such that its $E_p$ norm is
finite we can choose $\eps$ so that the right hand side of the above
inequality is negative, and so we conclude that the 
$E_p$ norm of $u$ at future times can not become larger that its
initial  value.
Furthermore if the function $F(\cdot)$ can be chosen to approach zero
at least linearly then for small enough $\eps$ there is an exponential
decay. 
But the above mentioned condition in 
$F(\cdot)$
follows directly from the smoothness assumptions on the coefficients
of the equations system, thus we have:

\begin{theorem}

Given any initial data $u_0$ whose $E_p$ norm is finite, then there exists $\eps_0 > 0$
such that for all $0 \leq \eps \leq \eps_0$ a solution exists and it decays exponentially to
zero.

\end{theorem}

For the case at hand one must be a bit more careful for the frame
chosen is not surface orthogonal, and furthermore the matrices $A^a_0$
are  not time independent. 
To control the first problem we shall use the background homogeneity
hypersurfaces for which the zeroth component of the background frame is 
surface-orthogonal, thus the contributions due to the
non-surface-orthogonality of the frame to the energy difference will be
of order $\eps$. 
We call that time $\tau$ ($=t_0$), and correspondingly choose $n_a := {}_0e^0_a$.
We also introduce the matrix $A^i$ so that the principal part becomes

\begin{equation}
  \label{eq:syst_hat}
  u_t = (A^i_0 + \eps A^i_1(u))e_i(u) + (\hat B_0 + \eps \hat B_1(u))u
\end{equation}
For the case at hand, namely the flat case, this change is trivial, and in particular $\hat B_0 = B_0$.

Defining now the energy four-vector as

\begin{equation}
  \label{eq:energy_vector2}
  \hat E^a := e^a_0 <u,Hu> + e^a_i<u,HA^i u> := e^a_0 \hat E^0 + e^a_i \hat E^i,
\end{equation}
we get a similar expression for eqn. (\ref{eq:div}),

\begin{eqnarray}
  \label{eq:div_hat}
  \nabla_a \hat{E}^a & = & <u_t,Hu> + <u,H u_t> + <u,H_t u> \\
          & + & <u,HA^i e_i(u)> + <u,e_i(HA^i)u> + <e_i(u), HA^i u>  \nn \\
          & + & \hat E^0 \nabla_a e^a_0 + \hat E^i \nabla_a e^a_i \nn \\
          & = & 2\Re <u,H(u_t + A^ie_i(u))> + <u,(H_t + e_i(HA^i))u> \nn \\
          & + & \hat E^0 \chi + \hat E^i (\Gamma^j_{ji} - a_i) \nn \\
          & = & 2\Re <u,H_0 B_0u>  + \hat E^0 {}_0\chi 
          + \eps [2\Re <u,(H_0+\eps H_1) B_1u> \nn \\
          & + & <u,((H_1)_t + e_i(H_0A^i_1+H_1 A^i_0 + \eps H_1A^i_1))u> 
          - \hat E^i ({}_1\Gamma^j_{ji} - {}_1a_i)] \nn 
\end{eqnarray}
where we have used that ${}_0a_i=0$, and for the flat case ${}_0\Gamma^j_{ki} = 0$. 
The only new zeroth order
term is $\hat E^0 {}_0 \chi = 3h_0 \hat E^0$. 
Since,
$\partial_{\mu}(a_0^{-3}\sqrt{-g} \hat E^{\mu}) = a_0^{-3}\sqrt{-g} \nabla_a \hat E^a - a_0^{-3}\sqrt{-g}\;{}_0\chi \hat E^0$, we get

\begin{eqnarray}
  \label{eq:partial_div_hat}
 \frac{1}{a_0^{-3}\sqrt{-g}}&&\!\!\!\!\!\!\!\!\!\!\partial_{\mu}(a_0^{-3}\sqrt{-g} \hat E^{\mu})  \\
               =&& <u_t,Hu> + <u,H u_t> + <u,H_t u> + <u,HA^i e_i(u)> \nn \\
               &&+  <u,e_i(HA^i)u> + <e_i(u), HA^i u> 
               + \hat E^i \nabla_a e^a_i \nn \\ 
               =&& 2\Re <u,H(u_t + A^i e_i(u))> 
               + <u,(H_t + e_i(HA^i))u> \nn \\ 
               &&+ \hat E^i (\Gamma^j_{ji} - a_i) \nn \\
               =&& 2\Re <u,H_0 B_0u>  + \eps [2\Re <u,(H_0+\eps H_1) B_1u> \nn \\
               &&+ <u,((H_1)_t + e_i(H_0A^i_1+H_1 A^i_0 + \eps H_1A^i_1))u> 
                     - \hat E^i ({}_1\Gamma^j_{ji} - {}_1a_i)] \nn 
\end{eqnarray}

Thus applying Gauss theorem to the coordinate divergence in the above
expression, and noticing that $a_0^{-3}\sqrt{-g} = 1 + O(\eps)$ we get
the desired energy estimated as for the four dimensional flat case
treated above. Notice that the new volume element is time independent
to zeroth order, and so the constant on the Sobolev embedding and on
other estimates can be taken to be also time independent.

\begin{theorem}

If the equation of state of the fluid is such that conditions (\ref{cond_alpha}-\ref{cond_nu}) 
are satisfied, then
given any initial data perturbation of an expanding flat homogeneous cosmology, $u_0$ whose 
$E_p$ norm is finite, there exists $\eps_0 > 0$
such that for all $0 \leq \eps \leq \eps_0$ a solution exists and it decays exponentially to
zero.

\end{theorem}


\subsection{A variant for the case in which $e^{\mu}_i$ is not in $L^2$.}
\label{subsec:variant}


In some cases it is not possible to have perturbations satisfying the
constraint equations and having the $L^2$ norm of $e^{\mu}_i$ finite,
this is because in the open ($R^3$) flat background case there could 
arise the need of considering massive perturbations, and they do not
decay at infinity sufficiently fast.  
In that case we can only have, $D_ae^{\mu}_i \in L^2$.  
There are several ways to deal with this
problem, and perhaps the best treatment is the one
in~\cite{CHOQUET.B:CauchyDM}. 
For the present problem a machinery such
as in~\cite{CHOQUET.B:CauchyDM} is not necessary, for the decay
properties can be established locally (pointwise) for part of the
equations.  The idea is not to look at the Sobolev norms of
$e^{\mu}_i$, but rather directly at its maximum norm of it, this is
possible because the time derivative of the frame does not have any
space derivative, it is just a ordinary differential equation. Thus,
if we call by $e$ the frame variables, and by $u$ all the others we
have a system of the form,

\begin{eqnarray}
  \label{eq:e-u}
  e_t & = & -h_0 e_t + f_u(u) + \eps f_e(e,u)  \\
  u_t & = & (A^a_0 + \eps A^a_1(e,u))D_a u + (B_0 + \eps B_1(u))u \nn
\end{eqnarray}
where here the matrices $A^a$, and $B$ are not the same as before, but are the corresponding restrictions.
The usual theory of energy bounds using the Gagliardo-Niremberg-Moser estimates, (see for instance 
{\bf Proposition 3.7}, and note bellow it, of  \cite{TaylorIII}) give us for the second equation, and the
corresponding equations for the space derivatives of both, $e$, and $u$, the following inequality,

\begin{eqnarray}
  \label{eq:energy_norm(u)}
 \lefteqn{\frac{d}{d\tau} (E_p(u) + E_{p-1}(De)) \leq } \\
       && -\delta (E_p(u) + E_{p-1}(De)) +
       \eps F(E_p(u), E_{p-1}(De), ||e||_{C^0} ), 
\end{eqnarray}
for some smooth $F$ with $F(0,\cdot, \cdot)=0$, since the negativity of the restricted $B_0$ as well as
for the whole $B_0$ (needed for the estimates on the system of space derivatives $(De,Du)$) follows as before.
Note that the $||De||_{C^0}$ is bounded by the $E_{p-1}(De)$ norm if 
$p  \geq 3$.

While for the first equation we get, at each point $q$ of the base manifold,

\begin{eqnarray}
  \label{eq:absol_e}
  |e(q,t)|_t & \leq & -\delta|e(q,t)| + G_u(E_p(u)) + \eps G_e(E_p(u), |e(q,t)|), 
\end{eqnarray}
With $G_u$, and $G_e$ smooth, polynomially bounded, functions of all its arguments,
and with $G_u(0) = 0$, and $G_e(\cdot, 0) = 0$.
In fact in our case $G_u$ is a linear function of the $E_p(u)$ norm.

To show exponential decay for $\eps$ small enough we assume, for contradiction, that given initial data
$(e_0, u_0)$ at $t=0$ there exists a time $T^{\star} > 0$ given by 

\begin{equation}
  \label{eq:T_star}
  T^{\star} = \inf_{T>0} :\left\{
  \begin{array}[c]{lll}
     E_p(u(T)) &=& E_p(u(0))(1+\Delta^2),\;\; or  \\
     E_{p-1}(De(T))&=& E_{p-1}(De(0))(1+\Delta^2),\;\; or \\
     ||e||_{C^0(T)} &=& ||e||_{C^0(0)} + \frac{C(E_p(u(0)))}{\delta} \}
  \end{array}
\right\}
\end{equation}
where $C(E_p(u))$ is a function
to be defined bellow, and $\Delta$ is some non-zero constant. 
We shall show that $T^{\star} = \infty$, and from the proof it also
follows that the solution not only exists for all times, but in fact
decays exponentially. 

Given the initial data we choose $\eps_G$ small enough so that 

\begin{eqnarray}
\lefteqn{\eps_G F(E_p(u(0))(1+\Delta^2), E_{p-1}(De(0))(1+\Delta^2),
  ||e||_{C^0(0)} + \frac{C(E_p(u(0)))}{\delta}) \leq } 
  && \hspace{7cm} \\
  && \hspace{7cm}
  \frac{\delta}{2}(E_p(u(0)) + E_{p-1}(De(0))) \nn
\end{eqnarray}
then it is clear that $(u,De)$ will decay as 
$e^{-\frac{\delta t}{2}}$ 
as long as $T \leq T^{\star}$

We now turn to the frame equation and choose $\eps_L$ so that
 
\begin{eqnarray}
\lefteqn{\eps_L G_e(E_p(u(0))(1+\Delta^2), ||e||_{C^0(0)} +
  \frac{C(E_p(u(0)))}{\delta}) \leq} &&
  \hspace{7cm} \\
  && \hspace{7cm}
  \frac{\delta}{2}(||e||_{C^0(0)} + \frac{C(E_p(u(0)))}{\delta}),
  \hspace{0cm} \nn
\end{eqnarray}
then we have, as long as $T \leq T^{\star}$

\begin{eqnarray}
  \label{eq:absol_e2}
  |e(q,t)|_t & \leq & -\frac{\delta}{2}|e(q,t)| + G_u(E_p(u)), 
\end{eqnarray}
and from this it follows that 

\begin{equation}
  \label{eq:exp_bound_e}
  ||e||_{C^0(T)} \leq (||e||_{C^0(0)} + \frac{\tilde{C}}{\delta}) e^{-\frac{\delta t}{2}},
\end{equation}
where $\tilde{C} = G_u(E_p(u))$. 
Thus we see that taking 
$C = \tilde{C}$, and $\eps = \min \{\eps_G, \eps_L \}$ 
we obtain that 
$T^{\star} = \infty$, 
and so the solution exists for all times and furthermore decays 
exponentially.  
Thus we have proven a similar Theorem as the one above for the case
where  the $L^2$ norm of the frame is not bounded, but rather its
$C^0$ norm is.

{\bf Acknowledgements:}

I thank H. Friedrich for pointing to me the possibility of using his description of the Einstein-Euler
system for this particular problem, and for several enlightening discussions. I also thank Uwe Brauer and 
Gabriel Nagy for pointing to aspects of earlier versions of this work which needed further refinements or
corrections.


\end{document}